\documentclass[12pt,reqno]{article}
\usepackage{amsfonts}
\usepackage{amsmath}
\usepackage{amsbsy}

\usepackage{amssymb,latexsym}
\numberwithin{equation}{section}

\begin{document}

\title{The Non-Split Scalar Coset in Supergravity Theories}
\author{Nejat T. Y$\i$lmaz\\
          Department of Physics\\
          Middle East Technical University\\
          06531 Ankara,Turkey\\
          \texttt{ntyilmaz@metu.edu.tr}}
\maketitle
\begin{abstract}
The general non-split scalar coset of supergravity theories is
discussed.The symmetric space sigma model is studied in two
equivalent formulations and for different coset
parametrizations.The dualisation and the local first order
formulation is performed for the non-split scalar coset $G/K$ when
the rigid symmetry group $G$ is a real form of a non-compact
semisimple Lie group (not necessarily split) and the local
symmetry group $K$ is $G$'s maximal compact subgroup.A comparison
with the scalar cosets arising in the $T^{10-D}$-compactification
of the heterotic string theory in ten dimensions is also
mentioned.

\end{abstract}

\section{Introduction}

The first order formulation of the maximal supergravities
($D\leq$11) has been given in [1,2] where the scalar sectors which
are governed by the $G/K$ symmetric space sigma model are studied
case by case.Dualisation of the fields and the generators which
parametrize the coset is the method used to obtain locally the
first order equations as a twisted self duality condition in
[1,2].In [3] a general formulation based on the structure
constants is given to construct an abstract method of dualisation
and to derive the first order equations for the $G/K$ coset sigma
model when the rigid symmetry group $G$ is in split real form
(maximally non-compact).Although the maximal supergravities fall
into this class there are more general cases of pure or matter
coupled supergarvity theories whose scalar sectors posses a
non-split rigid symmetry group.When the Bosonic sector of the ten
dimensional simple supergravity coupled to $N$ Abelian gauge
multiplets is compactified on $T^{10-D}$ the scalar sector of the
$D$-dimensional reduced theory can be formulated as $G/K$
symmetric space sigma model with $G$ being in general non-split
(not maximally non-compact) [4].In particular when $N$ is chosen
to be $16$ the formulation and the coset realizations of [4]
correspond to the $D$-dimensional Kaluza-Klein reduction of the
low-energy effective heterotic string theory in ten dimensions.

In this work by presenting an algebraic outline we are enlarging
the first order formulation of [3] by using the solvable algebra
parametrization [5,6] to a more general case which contains a
rigid symmetry group $G$ which is a real form of a non-compact
semisimple Lie group and $G$ is not necessarily in split real
form.We will choose a different spacetime signature convention
than the one assumed in [3].We will assume $s=1$ whereas in [3] it
has been taken as $s=(D-1)$.Therefore there will be a sign factor
depending on the spacetime dimension in the first-order
equations.The formalism presented here covers the split symmetry
group case as a particular limit in the possible choice of the
non-compact rigid symmetry groups as it will be clear in section
two.Before giving the first order formulation we will present two
equivalent definitions of the $G/K$ scalar coset sigma model and
we will study them in detail.We will introduce two coset maps,one
being the parametrization used in [1,2,3].A transformation law
between the two sets of scalar fields will also be established to
be used to relate the corresponding first order equations.The
second order equations of the vielbein formulation and the
internal metric formulation will be derived for both of these
coset maps.We will also discuss the correspondence of our
formulation with the scalar manifold cosets arising in [4] by
identifying the generators introduced in [4] and by inspecting the
coset parametrizations.

In section two we will introduce the algebraic outline of the
symmetric space sigma model.The Cartan and the Iwasawa
decompositions will be discussed.Section three is reserved for the
two equivalent formulations of the symmetric space sigma model.In
section four we will present the dualisation by doubling the
fields and the generators,we will also discuss the local first
order equations which are generalizations of the results of [3]
for two different parametrizations of the coset $G/K$.We will also
show that the same equations may be achieved by the dualisation
performed on the coset map which is different than the one used in
[1,2,3].Finally in section five we will mention about the
comparison of our construction with the coset realizations of [4].

\section{The Scalar Coset Manifold}

The scalar manifolds of all the pure and the matter coupled $N>2$
extended supergravities in $D=4,5,6,7,8,9$ dimensions as well as
the maximally extended supergravities in $D\leq11$ are homogeneous
spaces in the sense that they allow a transitive action of a Lie
group on them.They are in the form of a coset manifold $G/K$ where
$G$ is a real form of a non-compact semisimple Lie group and $K$
is a maximal compact subgroup of $G$.

For a real semisimple Lie algebra $\mathbf{g}_{0}$ each maximal
compactly imbedded subalgebra $\mathbf{k}_{0}$ corresponding to a
maximal compact subgroup of the Lie group of $\mathbf{g}_{0}$ is
an element of a Cartan decomposition [7]

\begin{equation}
\mathbf{g}_{0}=\mathbf{k}_{0}\oplus\mathbf{u}_{0}
\end{equation}

 which is a vector space direct sum such that if $\mathbf{g}$ is
the complexification of $\mathbf{g}_{0}$ and $\sigma$ is the
conjugation of $\mathbf{g}$ with respect to $\mathbf{g}_{0}$ then
there exists a compact real form $\mathbf{g}_{k}$ of $\mathbf{g}$
such that

\begin{equation}
\sigma(\mathbf{g}_{k})\subset\mathbf{g}_{k}\quad,\quad
\mathbf{k}_{0}=\mathbf{g}_{0}\cap\mathbf{g}_{k}\quad,\quad
\mathbf{u}_{0}=\mathbf{g}_{0}\cap(i\mathbf{g}_{k}).
\end{equation}

The proof of the existence of Cartan decompositions can be found
in [7].If we define a map $\omega :
\mathbf{g}_{0}\rightarrow\mathbf{g}_{0}$ such that $\omega(T +
X)=T - X \quad(\forall T \in\mathbf{k}_{0}$ and $\forall
X\in\mathbf{u}_{0})$ then ($\mathbf{g}_{0},\omega$) is an
orthogonal symmetric Lie algebra of non-compact type.The pair
($G,K$) associated with ($\mathbf{g}_{0},\omega$) is a Riemannian
symmetric pair.Therefore $K$ is a closed subgroup of $G$ and the
quotient topology on $G/K$ induced by $G$ generates a unique
analytical manifold structure and $G/K$ is a Riemannian globally
symmetric space for all the possible $G$-invariant Riemann
structures on $G/K$.The exponential map $Exp$ [7] is a
diffeomorphism from $\mathbf{u}_{0}$ onto the space
$G/K$.Therefore there is a legitimate parametrization of the coset
manifold by using $\mathbf{u}_{0}$.If $\{T_{i}\}$ are the basis
vectors of $\mathbf{u}_{0}$ and $\{\varphi^{i}(x)\}$ are
$C^{\infty}$ maps over the $D$-dimensional spacetime the map

\begin{equation}
  \nu(x)=e^{\varphi^{i}(x)T_{i}}
\end{equation}

is an onto $C^{\infty}$ map from the $D$-dimensional spacetime to
the Riemannian  globally symmetric space $G/K$.

If we define the Cartan involution $\theta$ as the involutive
automorphism of $\mathbf{g}_{0}$ for which the bilinear form
$-<X,\theta Y>$ is positive definite $\forall
X,Y\in\mathbf{g}_{0}$ then the root space of the Cartan subalgebra
(the maximal Abelian subalgebra) $\mathbf{h}_{0}$ of
$\mathbf{g}_{0}$ decomposes into two orthogonal components with
eigenvalues $\pm1$ under $\theta$ [6].If we denote the set of
invariant roots as $\Delta_{c}$ ($\theta(\alpha)=\alpha$) then
their intersection with the set of positive roots will be denoted
as $\Delta_{c}^{+}=\Delta^{+}\cap\Delta_{c}$.The remaining roots
in $\Delta^{+}$ namely $\Delta_{nc}^{+}=\Delta^{+}-\Delta_{c}^{+}$
are the ones whose corresponding generators $\{E_{\alpha}\}$ do
not commute with the elements of
$\mathbf{h}_{k}=\mathbf{h}_{0}\cap\mathbf{u}_{0}$ where
$\mathbf{h}_{k}$ is the maximal Abelian subspace of
$\mathbf{u}_{0}$ and it consists of the non-compact part of
$\mathbf{h}_{0}$ [5].

The Cartan subalgebra generates an Abelian subgroup in $G$ which
is called the torus [6].Although we call it torus it is not the
ordinary torus topologically in fact it has the topology
$(S^{1})^{m}\times \mathbb{R}^{n}$ for some $m$ and $n$ and if it
is diagonalizable in $\mathbb{R}$ (such that $m=0$) then it is
called an $R$-split torus.These definitions can be generalized for
the subalgebras of $\mathbf{h}_{0}$ aswell.The subspace of $G$
which is generated by $\mathbf{h}_{k}$ is the maximal $R$-split
torus in $G$ in the sense defined above and it's dimension is
called the $R$-rank which we will denote by $r$.There are two
special classes of semisimple real forms.If $r$ is maximal such
that $r=l$ where $l$ is the rank of $G$
($l=$dim$(\mathbf{h}_{0})$),which also means
$\mathbf{h}_{k}=\mathbf{h}_{0}$ then the Lie group $G$ is said to
be in split real form (maximally non-compact).If on the other hand
$r$ is minimal such that $r=0$ then $G$ is a compact real form.All
the other cases in between are non-compact semisimple real forms.

Since we assume that $G$ is a non-compact real form of a
semisimple Lie group the Iwasawa decomposition [7] enables a link
between the Cartan decomposition and the root space decomposition
of the semisimple real lie algebra $\mathbf{g}_{0}$

\begin{equation}
 \mathbf{g}_{0}=\mathbf{k}_{0}\oplus\mathbf{s}_{0}
\end{equation}

where $\mathbf{s}_{0}$ is a solvable Lie algebra of
$\mathbf{g}_{0}$ and the sum is a direct vector space sum like in
(2.1).A Lie algebra $\mathbf{g}$ is called solvable if it's n-th
derived algebra is $\{0\}$ denoted as $D^{n}\mathbf{g}={0}$ where
the first derived algebra is the ideal of $\mathbf{g}$ generated
by the commutator $[X,Y]$ of all of it's elements $X,Y$ and the
higher order derived algebras are defined inductively in the same
way over one less rank derived algebra.Starting from the Cartan
decomposition one can show that $\mathbf{s}_{0}$ indeed exists and
Iwasawa decomposition is possible for a non-compact real form of a
semisimple Lie group [5,7].

The real Lie algebra $\mathbf{g}_{0}$ is a real form of  the
complex Lie algebra $\mathbf{g}$.We will denote the complex Lie
subalgebra of $\mathbf{g}$ generated by the positive root
generators $\{E_{\alpha}\}$ for $\alpha\in\Delta_{nc}^{+}$ as
$\mathbf{n}$.A real Lie subalgebra of $\mathbf{g}_{0}$ can then be
defined as $\mathbf{n}_{0}=\mathbf{g}_{0}\cap\mathbf{n}$.Both
$\mathbf{n}$ and $\mathbf{n}_{0}$ are nilpotent Lie algebras.The
solvable real Lie algebra $\mathbf{s}_{0}$ of $\mathbf{g}_{0}$
defined in the Iwasawa decomposition in (2.4) can then be written
as a direct vector space sum [7]

\begin{equation}
 \mathbf{s}_{0}=\mathbf{h}_{k}\oplus\mathbf{n}_{0}
\end{equation}

where $\mathbf{h}_{k}=\mathbf{h}_{0}\cap\mathbf{u}_{0}$ as defined
before.Therefore the elements of the coset $G/K$ are in one-to-one
correspondence with the elements of $\mathbf{s}_{0}$ trough the
exponential map as a result of (2.4).This parametrization of the
coset $G/K$ is called the solvable Lie algebra parametrization
[5].

\section{The Sigma Model}

We will now present two equivalent formulations of the symmetric
space sigma model which governs the scalar sector of a class of
supergravity theories which have homogeneous coset scalar
manifolds as mentioned in the previous section.The first of these
formulations does not specify a coset parametrization while the
second one makes use of the results of the Iwasawa
decomposition.The first formulation is a more general one which is
valid for any Lie group $G$ and it's subgroup $K$ namely it is the
$G/K$ non-linear sigma model.In particular it is applicable to the
symmetric space sigma models.If we consider the set of $G$-valued
maps $\nu(x)$ which transform onto each other as $\nu\rightarrow
g\nu k(x)\quad\forall g\in G,k(x)\in K$ we can calculate
$\mathcal{G}=\nu^{-1}d\nu$ which is the pull back of the Lie
algebra $\mathbf{g}_{0}$-valued Cartan form over $G$ through the
map $\nu(x)$.The map $\nu(x)$ can always be chosen to be a
parametrization of the coset $G/K$.Moreover if $G$ obeys the
conditions presented in the last section $\nu(x)$ can be taken as
the map (2.3) $\nu(x)=e^{\varphi^{i}(x)T_{i}}$ by using the Cartan
decomposition.As it is clear from section two when $G$ is a real
form of a non-compact semisimple Lie group
 we can function the Iwasawa decomposition resulting in the solvable Lie algebra parametrization
 aswell.For the most general case of $\nu(x)$ we have

\begin{equation}
\mathcal{G}_{\mu}dx^{\mu}=(f_{\mu}^{a}(x)T_{a}+\omega_{\mu}^{i}(x)K_{i})dx^{\mu}
\end{equation}

where $T_{a}\in\mathbf{u}_{0}$ and $K_{i}\in\mathbf{k}_{0}$.Here
$\mathbf{u}_{0}$ is the orthogonal complement of $\mathbf{k}_{0}$
in $\mathbf{g}_{0}$.In particular if $G$ is a real form of a
semisimple Lie group and $K$ is a maximal compact subgroup of $G$
then the Cartan decomposition (2.1) can be used.When $G/K$ is a
Riemannian globally symmetric space then the fields
$\{f_{\mu}^{a}\}$ form a vielbein of the $G$-invariant Riemann
structures on $G/K$ and $\{\omega_{\mu}^{i}\}$ can be considered
as the components of the connection one forms of the gauge theory
over the $K$-bundle.We should also bear in mind that
$[\mathbf{k}_{0},\mathbf{k}_{0}]\subset\mathbf{k}_{0}$ and if
furthermore $[\mathbf{k}_{0},\mathbf{u}_{0}]\subset\mathbf{u}_{0}$
then we will have a simpler theory.Let $P_{\mu}\equiv
f_{\mu}^{a}T_{a}$ and $Q_{\mu}\equiv\omega_{\mu}^{i}K_{i}$ then we
can construct a Lagrangian [8,9]

\begin{equation}
\mathcal{L}=\frac{1}{2}tr(P_{\mu}P^{\mu})
\end{equation}

where the trace is over the representation chosen.$\mathcal{L}$ is
invariant when $\nu(x)$ is transformed under the rigid (global)
action of $G$ from the left and the local action of $K$ from the
right as given above.The elements of (3.1) $P_{\mu}$ and $Q_{\mu}$
are invariant under the rigid action of $G$ but under the local
action of $K$ they transform as

\begin{equation}
\begin{aligned}
Q_{\mu}&\rightarrow k(x)Q_{\mu}k^{-1}(x)+k(x)\partial_{\mu} k^{-1}(x)\\
\\
P_{\mu}&\rightarrow k(x)P_{\mu}k^{-1}(x).
\end{aligned}
\end{equation}

The field equations corresponding to (3.2) are

\begin{equation}
\begin{aligned}
D_{\mu}P^{\mu}&=\partial_{\mu}P^{\mu}+[Q_{\mu},P^{\mu}]\\
\\
&=0
\end{aligned}
\end{equation}

where we have introduced the covariant derivative
$D_{\mu}\equiv\partial_{\mu}+[Q_{\mu},\quad]$.

Besides having more general applications the above formalism
covers the symmetric space sigma model in it.We will now introduce
another parametrization of the coset $G/K$ which is locally
legitimate [7] as a result of the solvable Lie algebra
parametrization of the Iwasawa decomposition which is discussed in
the last section.When $G$ is a real form of a non-compact
semisimple Lie group and $K$ is it's maximal compact subgroup the
coset $G/K$ can locally be parametrized as

\begin{equation}
\begin{aligned}
\nu (x)&=\mathbf{g}_{H}(x)\mathbf{g}_{N}(x)\\
\\
&=e^{\frac{1}{2}\phi ^{i}(x)H_{i}}e^{\chi ^{m}(x)E_{m}}
\end{aligned}
\end{equation}

where $\{H_{i}\}$ is a basis for $\mathbf{h}_{k}$ for $i=1,...,r$
and $\{E_{m}\}$ is a basis for $\mathbf{n}_{0}$.The map (3.5) is
obtained by considering a map from a coordinate chart of the
spacetime onto a neighborhood of the identity of
$\mathbf{s}_{0}$.Thus it defines locally a diffeomorphism from
$\mathbf{h}_{k}\times\mathbf{n}_{0}$ into the space $G/K$ so it is
a local parametrization of $G/K$.We assume that the locality is
both over the spacetime and over the space
$\mathbf{h}_{k}\times\mathbf{n}_{0}$ in order to write two
products of exponentials instead of the solvable Lie algebra
parametrization.If $G$ is in split real form then
$\mathbf{h}_{k}=\mathbf{h}_{0}$ and $\Delta_{nc}^{+}=\Delta^{+}$
thus the solvable algebra $\mathbf{s}_{0}$ becomes the Borel
subalgebra of $\mathbf{g}_{0}$.The fields $\{\phi^{i}\}$ are
called the dilatons and $\{\chi^{m}\}$ are called the axions.At
this stage we can calculate the field equations (3.4) in terms of
these newly defined fields under  the parametrization (3.5).The
Cartan form $\mathcal{G}=\nu^{-1}d\nu$ can be calculated from
(3.5) as follows

\begin{equation}
\begin{aligned}
\mathcal{G}&=\nu ^{-1} d\nu\\
\\
&=(\mathbf{g}_{N}^{-1}\mathbf{g}_{H}^{-1})(d\mathbf{g}_{H}\mathbf{g}_{N}+\mathbf{g}_{H}d\mathbf{g}%
_{N})\\
\\
&=\mathbf{g}_{N}^{-1}
d\mathbf{g}_{N}+\mathbf{g}_{N}^{-1}\mathbf{g}_{H}^{-1}d\mathbf{g}_{H}\mathbf{g}_{N}.
\end{aligned}
\end{equation}

If we make use of the identity $e^{-C}
de^{C}=dC-\frac{1}{2!}[C,dC]+\frac{1}{3!}[C,[C, dC]]-....$ in a
matrix representation the first term can be calculated as

\begin{equation}
\begin{aligned}
\mathbf{g}_{N}^{-1}d\mathbf{g}_{N}&=e^{-\chi ^{m}E_{m}} de^{\chi ^{m}E_{m}}\\
\\
&=d\chi ^{m}E_{m}-\frac{1}{2!}[\chi ^{m}E_{m},d%
\chi ^{n}E_{n}]+\frac{1}{3!}[\chi ^{m}E_{m},[\chi ^{l}E_{l},d\chi
^{n}E_{n}]]-....\\
\\
&=d\chi ^{m}E_{m}-\frac{1}{2!}\chi ^{m}d\chi
^{n}K_{mn}^{v}E_{v}+\frac{1}{3!}\chi ^{m}\chi ^{l}d\chi
^{n}K_{ln}^{v}K_{mv}^{u}E_{u}-....\\
\\
&=\overset{\rightharpoonup}{\mathbf{E }}\:\mathbf{\Sigma
}\:\overset{\rightharpoonup }{d\chi }
\end{aligned}
\end{equation}

where we have defined the row vector
$(\overset{\rightharpoonup}{\mathbf{E }})_{\alpha}=E_{\alpha}$ and
the column vector $(\overset{\rightharpoonup }{d\chi})^{\alpha}
=d\chi^{\alpha}$.We have also introduced $\mathbf{\Sigma }$ as the
dim$\mathbf{n}_{0}\times$dim$\mathbf{n}_{0}$ matrix

\begin{equation}
 \mathbf{\Sigma}=\sum\limits_{n=0}^{\infty }\dfrac{(-1)^{n}\omega
^{n}}{(n+1)!}
\end{equation}

$\omega _{\beta }^{\gamma }=\chi ^{\alpha }\,K_{\alpha \beta
}^{\gamma }$ where the structure constants $K_{\alpha \beta
}^{\gamma }$ are defined as $[E_{\alpha },E_{\beta }]=K_{\alpha
\beta }^{\gamma }\,E_{\gamma }$.If we consider the commutator
$[E_{\alpha },E_{\beta }]=N_{\alpha ,\beta }E_{\alpha +\beta }$
then $K_{\beta \beta }^{\alpha }=0$ also $K_{\beta \gamma }^{\alpha }=N_{\beta ,\gamma }$ if in the root sense $%
\beta +\gamma =\alpha $ and $K_{\beta \gamma }^{\alpha }=0$ if
$\beta +\gamma \neq \alpha $.Similarly since the Cartan generators
commute with each other we have

\begin{equation}
\begin{aligned}
\mathbf{g}_{H}^{-1} d\mathbf{g}_{H}&= e^{-\frac{1}{2}\phi
^{i}H_{i}} de^{\frac{1}{2}\phi
^{i}H_{i}}\\
\\
&=\frac{1}{2}d\phi ^{i}H_{i}.
\end{aligned}
\end{equation}

The second term in (3.6) can now be calculated as

\begin{equation}
\begin{aligned}
\mathbf{g}_{N}^{-1}(\frac{1}{2}d\phi ^{i}H_{i})\mathbf{g}_{N}&=%
e^{-\chi^{m}E_{m}}(\frac{1}{2}d\phi ^{i}H_{i})e^{\chi^{m}E_{m}}\\
\\
&=\frac{1}{2}d\phi ^{i}H_{i}-[\chi ^{m}E_{m},\frac{1}{2}d\phi
^{i}H_{i}]\\
\\
&+\frac{1}{2!}[\chi
^{m}E_{m},[\chi ^{l}E_{l},\frac{1}{2}d\phi ^{i}H_{i}]]-....\\
\\
&=\frac{1}{2}d\phi ^{i}H_{i}+\chi ^{m}\frac{1}{2}d\phi
^{i}m_{i}E_{m}\\
\\
&-\frac{1}{2!}\chi ^{m}\chi ^{l}\frac{1}{2}d\phi ^{i}l_{i}K_{ml}^{u}E_{u}+....\\
\\
&=\frac{1}{2}d\phi ^{i}H_{i}+\overset{\rightharpoonup}{\mathbf{E
}}\:\mathbf{\Sigma }\:\overset{\rightharpoonup }{U}
\end{aligned}
\end{equation}

where we have used the Campbell-Hausdorff formula
$e^{-X}Ye^{X}=Y-[X,Y]+\frac{1}{2!}[X,[X,Y]]-....$ and we have
defined the column vector $(\overset{\rightharpoonup
}{U})^{m}=\frac{1}{2}\chi ^{m}m_{i} d\phi ^{i}$.Also we have
$[H_{i},E_{m}]=m_{i}E_{m}$.Therefore the Cartan form
$\mathcal{G}=\nu^{-1}d\nu$ in (3.6) becomes

\begin{equation}
\mathcal{G}=\frac{1}{2}d\phi
^{i}H_{i}+\overset{\rightharpoonup}{\mathbf{E }}\:\mathbf{\Sigma
}\:(\overset{\rightharpoonup }{U}+\overset{\rightharpoonup
}{d\chi}).
\end{equation}

Since the expansion of $\mathcal{G}$ consists of only the
generators of $\mathbf{s}_{0}$ but not the generators of
$\mathbf{k}_{0}$ which is a direct result of (3.5) where the
parametrization is derived locally from the solvable Lie algebra
parametrization we have $Q_{\mu}=0$ and from (3.11) $P_{\mu}$ is

\begin{equation}
P_{\mu}=\frac{1}{2}\partial_{\mu}\phi
^{i}H_{i}+\Sigma_{m}^{\alpha}(\frac{1}{2}\chi ^{m}m_{i}
\partial_{\mu}\phi ^{i}+\partial_{\mu}\chi^{m})E_{\alpha}.
\end{equation}

Since $Q_{\mu}=0$ from (3.4) the equations of motion become

\begin{equation}
\partial^{\mu}P_{\mu}=0.
\end{equation}

Thus we have

\begin{gather}
\partial^{\mu}\partial_{\mu}\phi
^{i}=0\notag\\
\notag\\
\partial^{\mu}\,(\Sigma_{m}^{\alpha}\,(\frac{1}{2}\chi ^{m}m_{i}
\partial_{\mu}\phi ^{i}+\partial_{\mu}\chi^{m}))=0
\end{gather}

for $i=1,...r$ and $\alpha\in\Delta_{nc}^{+}$.

Another formulation of the $G/K$ symmetric space sigma model (with
$G$ not necessarily split) can be done by introducing the internal
metric $\mathcal{M}$.In this formulation the Lagrangian which is
invariant under the global action of $G$ from the left and the
local action of $K$ from the right is [1,2,10]

\begin{equation}
 \mathcal{L}=\frac{1}{4}tr(\ast d\mathcal{M}^{-1}\wedge d\mathcal{M})
\end{equation}

where the internal metric $\mathcal{M}$ is defined as
$\mathcal{M=}\nu ^{\#}\nu $ and $\#$ is the generalized transpose
over the Lie group $G$ such that $(exp(g))^{\#}=exp(g^{\#})$.It is
induced by the Cartan involution $\theta$ over $\mathbf{g}_{0}$
($g^{\#}=-\theta(g)$) [1,10].The Lagrangian can be expressed as

\begin{equation}
 \mathcal{L=-}\frac{1}{2}tr(\ast d\nu \nu ^{-1}\wedge (d\nu \nu
^{-1})^{\#}+\ast d\nu \nu ^{-1}\wedge d\nu \nu ^{-1}).
\end{equation}

We will again assume the parametrization given in (3.5).By
following the same steps given in detail in [3] we can calculate
the $\mathbf{s}_{0}$-valued one form $\mathcal{G}_{0}=d\nu \nu
^{-1}$.This is possible because the Borel parametrization used in
[3] is a limit case of the solvable Lie algebra parametrization as
discussed in section two so as far as the commutation relations
are concerned the equivalence between the general case and the
split case of [3] is straightforward.Thus from (3.5) we have

\begin{equation}
\begin{aligned}
\mathcal{G}_{0}&=d\nu \nu ^{-1}\\
\\
&=\frac{1}{2}d\phi ^{i}H_{i}+e^{%
\frac{1}{2}\alpha _{i}\phi ^{i}}F^{\alpha }E_{\alpha }\\
\\
&=\frac{1}{2}d\phi ^{i}H_{i}+\overset{\rightharpoonup }{
\mathbf{E}^{\prime }}\:\mathbf{\Omega }\:\overset{\rightharpoonup
}{d\chi }.
\end{aligned}
\end{equation}

where $\{H_{i}\}$ for $i=1,...,r$ are the generators of
$\mathbf{h}_{k}$ and $\{E_{\alpha}\}$ for
$\alpha\in\Delta_{nc}^{+}$ are the generators of
$\mathbf{n}_{0}$.We have defined
$F^{\alpha}=\mathbf{\Omega}^{\alpha}_{\beta}d\chi^{\beta}$ and the
row vector $(\overset{\rightharpoonup }{
\mathbf{E}^{\prime }})_{\alpha}=e^{%
\frac{1}{2}\alpha _{i}\phi ^{i}}E_{\alpha}$.Also $\mathbf{\Omega}$
is a dim$\mathbf{n}_{0}\times$dim$\mathbf{n}_{0}$ matrix

\begin{equation}
\begin{aligned}
 \mathbf{\Omega}&=\sum\limits_{n=0}^{\infty }\dfrac{\omega
^{n}}{(n+1)!}\\
\\
&=(e^{\omega}-I)\,\omega^{-1}
\end{aligned}
\end{equation}

The matrix $\omega$ has been already defined before.The equations
of motion of the Lagrangian (3.15) can be found as [6,10]

\begin{equation}
\begin{aligned}
d(\ast d\phi ^{i})&=\frac{1}{2}\sum\limits_{\alpha\in\Delta_{nc}^{+}}^{}\alpha _{i}%
e^{\frac{1}{2}\alpha _{i}\phi ^{i}}F^{\alpha }\wedge e^{%
\frac{1}{2}\alpha _{i}\phi ^{i}}\ast F^{\alpha }\\
\\
d(e^{\frac{1}{2}\gamma _{i}\phi ^{i}}\ast F^{\gamma
})&=-\frac{1}{2}\gamma _{j}e^{\frac{1}{2}\gamma _{i}\phi
^{i}}d\phi ^{j}\wedge \ast F^{\gamma }\\
\\
&+\sum\limits_{\alpha -\beta =-\gamma }e^{\frac{1}{2} \alpha
_{i}\phi ^{i}}e^{\frac{1}{2}\beta _{i}\phi ^{i}}N_{\alpha ,-\beta
}F^{\alpha }\wedge \ast F^{\beta }
\end{aligned}
\end{equation}

where $i,j=1,...,r$ and $\alpha,\beta,\gamma\in\Delta_{nc}^{+}$.We
have put the second equation in a convenient  form for the
analysis we will use in the dualisation section as in [3].

We will now introduce a transformation between the two
parametrizations given in (2.3) and (3.5) which are based on two
different sets of scalar functions.We can derive a procedure to
calculate the transformation between these two sets.We may assume
that the coset valued maps in (2.3) and (3.5) can be chosen to be
equal.This is possible if we restrict the scalar maps with the
ones which generate ranges in sufficiently small neighborhoods
around the identity element of $\mathbf{g}_{0}$ when they are
coupled to the algebra generators in (2.3) and (3.5) [7].This
local equality is sufficient since our aim is to obtain the local
first order formulation of the parametrization of (2.3) from the
first order formulation which is based on (3.5) in the next
section.We will firstly show a method through which one can
calculate the exact transformations from $\{\varphi^{i}\}$ to
$\{\phi^{j},\chi^{m}\}$.We will not attempt to solve the explicit
transformation functions which are dependent on the structure
constants in a complicated way.One can solve these set of
differential equations when the structure constants are
specified.Let us first define the function

\begin{equation}
f(\lambda)=e^{\lambda(\frac{1}{2}\phi^{i}H_{i})}e^{\lambda(\chi^{\alpha}E_{\alpha})}.
\end{equation}

Taking the derivative of $f(\lambda)$ with respect to $\lambda$
gives

\begin{equation}
\frac{\partial
f(\lambda)}{\partial\lambda}f^{-1}(\lambda)=\frac{1}{2}\phi^{i}H_{i}+e^{\frac{\lambda}{2}
\phi^{i}\alpha_{i}}\chi^{\alpha}E_{\alpha}.
\end{equation}

We have  used the fact that $e^{\frac{\lambda}{2}
\phi^{i}H_{i}}\chi^{\alpha}E_{\alpha}e^{-\frac{\lambda}{2}
\phi^{i}H_{i}}=e^{\frac{\lambda}{2}
\phi^{i}\alpha_{i}}\chi^{\alpha}E_{\alpha}$.Now if we let
$f(\lambda)=e^{C(\lambda)}$ where we define
$C(\lambda)=\varphi^{i}(\lambda)T_{i}$ and use the formula
$de^{C}e^{-C}=dC+\frac{1}{2!}[C,dC]+\frac{1}{3!}[C,[C, dC]]+....$
we find that

\begin{equation}
\frac{1}{2}\phi^{i}H_{i}+e^{\frac{\lambda}{2}
\phi^{i}\alpha_{i}}\chi^{\alpha}E_{\alpha}=\overset{\rightharpoonup
}{ \mathbf{T}}\:\mathbf{S}(\lambda)\:\overset{\rightharpoonup }{
\partial\varphi}
\end{equation}

where the components of the row vector $\overset{\rightharpoonup
}{ \mathbf{T}}$ are $T_{i}=H_{i}$ for $i=1,...,r$ and
$T_{\alpha+r}=E_{\alpha}$ for
$\alpha=1,...,$dim$\mathbf{n}_{0}$.Besides the column vector
$\overset{\rightharpoonup }{\partial\varphi}$ is defined as
$\{\frac{\partial\varphi^{i}(\lambda)}{\partial\lambda}\}$.We have
also introduced the dim$\mathbf{s}_{0}\times$dim$\mathbf{s}_{0}$
matrix $\mathbf{S}(\lambda)$ as
\begin{equation}
\begin{aligned}
\mathbf{S}(\lambda)&=\sum\limits_{n=0}^{\infty }\dfrac{V
^{n}(\lambda)}{(n+1)!}\\
\\
&=(e^{V(\lambda)}-I)V^{-1}(\lambda)
\end{aligned}
\end{equation}

The matrix $V(\lambda)$ is
$V_{\alpha}^{\beta}(\lambda)=\varphi^{i}(\lambda)C_{i\alpha}^{\beta}$
for $[T_{i},T_{j}]=C_{ij}^{k}T_{k}$.The calculation of the right
hand side of (3.22) is similar to (3.17).If the structure
constants are given for a particular $\mathbf{s}_{0}$ one can
obtain the functions $\{\varphi^{i}(\lambda)\}$ from the set of
differential equations  (3.22).Then setting $\lambda=1$ will yield
the desired set of functions
$\{\varphi^{i}(\phi^{j},\chi^{\alpha})\}$.We might also make use
of a direct calculation namely the Lie's theorem [7].For a matrix
representation and in a neighborhood of the identity if we let
$e^{C}=e^{A}e^{B}$ then

\begin{equation}
\begin{aligned}
 C&=B+\int\limits_{0}^{1}g(e^{tadA}e^{adB})Adt\\
 \\
 &=A+B+\frac{1}{2}[A,B]+\frac{1}{12}([A,[A,B]]-[B,[B,A]])+...
 \end{aligned}
 \end{equation}
 where $g\equiv lnz/z-1$.In the above equation if we choose
 $A=\frac{1}{2}\phi^{i}H_{i}$,$B=\chi^{\alpha}E_{\alpha}$ and
 $C=\varphi^{i}T_{i}$ we can calculate the transformations we need.

 We may also derive the differential form of the transformation between
 the two parametrizations which is more essential for our purposes of obtaining the local first
 order formulation of the parametrization in (2.3) in the next section.From (2.3) by choosing $\{T_{i}\}$ as the basis of
 $\mathbf{s}_{0}$,similar to the previous calculations we can calculate the
 $\mathbf{s}_{0}$-valued Cartan form $d\nu\nu^{-1}$ as
 \begin{equation}
 \begin{aligned}
d\nu\nu^{-1}&=de^{\varphi^{i}T_{i}}e^{-\varphi^{i}T_{i}}\\
\\
&=\overset{\rightharpoonup }{
\mathbf{T}}\:\mathbf{\Delta}\:\overset{\rightharpoonup }{\mathbf{
d\varphi}}.
\end{aligned}
\end{equation}

We have defined $\overset{\rightharpoonup }{ \mathbf{T}}$ before
$\overset{\rightharpoonup }{\mathbf{d\varphi}}$ is a column vector
of the field strengths $\{d\varphi^{i}\}$ and the
dim$\mathbf{s}_{0}\times$dim$\mathbf{s}_{0}$ matrix
$\mathbf{\Delta}$ can be given as
\begin{equation}
\begin{aligned}
\mathbf{\Delta}&=\sum\limits_{n=0}^{\infty }\dfrac{M
^{n}}{(n+1)!}\\
\\
&=(e^{M}-I)M^{-1}
\end{aligned}
\end{equation}

 where $M_{\alpha}^{\beta}=\varphi^{i}C_{i\alpha}^{\beta}$.We should
imply that $\mathbf{\Delta}=\mathbf{S}(\lambda=1)$ and
$M=V(\lambda=1)$.If we refer to the equation (3.17) we have
already calculated the $\mathbf{s}_{0}$-valued Cartan form for the
parametrization of (3.5).Therefore if we compare (3.17) and (3.25)
since locally they must be equal we find
\begin{equation}
\begin{aligned}
\mathbf{\Delta}_{i}^{\gamma}d\varphi^{i}&=\frac{1}{2}d\phi^{\gamma}\\
\\
\mathbf{\Delta}_{i}^{\beta}d\varphi^{i}&=e^{\frac{1}{2}\beta_{j}\phi^{j}}\mathbf{\Omega}_{k}^{\beta}
d\chi^{k}.
\end{aligned}
\end{equation}

The indices above are $\gamma=1,...,r$;
$\beta,k=r+1,...,$dim$\mathbf{n}_{0}$; $
i=1,...,$dim$\mathbf{s}_{0} =r+$dim$\mathbf{n}_{0}$ and
$\beta\in\Delta_{nc}^{+}$.As a result we have obtained the
differential form of the transformation between $\{\varphi^{i}\}$
and $\{\phi^{j},\chi^{\alpha}\}$.It can be seen that the relation
between the two scalar parametrizations is dependent on the
structure constants in a very complicated way.One can also
integrate (3.27) to obtain the explicit form of this
transformation as an alternative to the equation (3.22).

Finally we will calculate the field equations (3.4) for the
parametrization (2.3) by assuming the Iwasawa
 decomposition.Smilar to (3.25) from (2.3) the $\mathbf{s}_{0}$-valued Cartan form
 $\nu^{-1}d\nu$ is
 \begin{equation}
 \begin{aligned}
\nu^{-1}d\nu&=e^{-\varphi^{i}T_{i}}de^{\varphi^{i}T_{i}}\\
\\
&=\overset{\rightharpoonup }{
\mathbf{T}}\:\mathbf{W}\:\overset{\rightharpoonup }{\mathbf{
d\varphi}}.
\end{aligned}
\end{equation}

where we have

\begin{equation}
\begin{aligned}
\mathbf{W}&=\sum\limits_{n=0}^{\infty }\dfrac{(-1)^{n}M
^{n}}{(n+1)!}\\
\\
&=(I-e^{-M})M^{-1}.
\end{aligned}
\end{equation}

From (3.28) like in (3.11) we see that $Q_{\mu}=0$ due to the
solvable Lie algebra parametrization.On the other hand $P_{\mu}$
is

\begin{equation}
\begin{aligned}
P_{\mu}&=P_{\mu}^{i}\,T_{i}\\
\\
&=(\mathbf{W})\,_{k}^{l}\,\partial_{\mu}\varphi^{k}\,T_{l}.
\end{aligned}
\end{equation}

Thus in terms of the fields $\{\varphi^{i}\}$ the second order
equations (3.4) become

\begin{equation}
\begin{aligned}
\partial^{\mu}\,P_{\mu}&=\partial^{\mu}\,((\mathbf{W})\,_{k}^{l}\,\partial_{\mu}\varphi^{k}T_{l})\\
\\
&=0.
\end{aligned}
\end{equation}

\section{Dualisation and the First Order Formulation}

The local first order formulation of the $G/K$ symmetric space
sigma model when $G$ is in split real form has been given in [3]
where we have applied the standard dualisation method of [1,2] by
introducing dual generators for the Borel subgroup generators and
also new auxiliary fields $(D-2)$-forms for the scalar fields.Then
the enlarged Lie superalgebra which contains the original Borel
algebra has been inspected so that it would realize the original
second order equations in an enlarged coset model.In [3] after
calculating the extra commutation relations coming from the new
generators locally the first order equations were given as a
twisted self-duality equation $\ast
\mathcal{G^{\prime}=SG^{\prime}}$ where $\mathcal{G^{\prime}}$ is
the doubled field strength (the Cartan form generated by the new
coset representative) and $\mathcal{S}$ is a pseudo-involution of
the enlarged Lie superalgebra which for the special case of the
scalar coset maps the original generators  onto the dual ones and
the dual generators onto the original scalar generators with a
sign factor depending on the dimension $D$ and the signature $s$
of the spacetime as explained in [1,2,3].The split rigid group
symmetric space sigma model is a limiting case as discussed in
section two.The solvable algebra is a subalgebra of the Borel
algebra in general and the field equations (3.19) for the general
non-compact (not necessarily split) real form model are in the
same form with the split case except the summing indices.Therefore
the results in [3] can be generalized for the general non-compact
symmetric space sigma model.We will give a summary of the results
which are generalizations  and whose detailed calculations are
similar to the ones in [3].

Firstly we will introduce dual $(D-2)$-form fields for the
dilatons and the axions which are defined in (3.5).The dual fields
will be denoted as $\{\widetilde{\phi^{i}}\}$ and
$\{\widetilde{\chi^{m}}\}$.For each scalar generator we will also
define dual generators which will extend the solvable Lie algebra
$\mathbf{s}_{0}$ to a Lie superalgebra which generates a
differential algebra with the local differential form algebra
[2].These generators are $\{\widetilde{E}_{m}\}$ as duals of
$\{E_{m}\}$ and $\{\widetilde{H}_{i}\}$ for $\{H_{i}\}$.If we
define a new parametrization into the enlarged group

\begin{equation}
\nu ^{\prime }(x)=e^{\frac{1}{2}\phi ^{i}H_{i}}e^{\chi
^{m}E_{m}}e^{\widetilde{\chi }^{m}\widetilde{E}_{m}}e^{
\frac{1}{2}\widetilde{\phi }^{i}\widetilde{H}_{i}}
\end{equation}

we can calculate the doubled field strength
$\mathcal{G}^{\prime}=d\nu ^{\prime }\nu ^{\prime -1}$ and then
use the twisted self-duality condition $\ast
\mathcal{G^{\prime}=SG^{\prime}}$ to find the structure constants
of the dual generators and the first order equations of motion.The
general form of the commutation relations in addition to the ones
of $\mathbf{s}_{0}$ can be given as [1,2]

\begin{gather}
[E_{\alpha },\widetilde{T}_{m}]=\widetilde{f}_{\alpha
m}^{n}\widetilde{T} _{n}\quad, \quad
[H_{i},\widetilde{T}_{m}]=\widetilde{g}_{im}^{n}
\widetilde{T}_{n},\notag\\
\notag\\
[\widetilde{T}_{m},\widetilde{T}_{n}]=0
\end{gather}

where $\widetilde{T}_{i}=\widetilde{H}_{i}$ for $i=1,...,r$ and
$\widetilde{T}_{\alpha+r}=\widetilde{E}_{\alpha}$ for
$\alpha=1,...,$dim$\mathbf{n}_{0}$.In general the
pseudo-involution $\mathcal{S}$ maps the original generators onto
the dual ones and it has the same eigenvalues $\pm1$ with the
action of the operator $\ast\circ\ast$ on the corresponding dual
field strength of the coupling dual potential.Therefore
$\mathcal{S}T_{i}=\widetilde{T}_{i}$ and
$\mathcal{S}\widetilde{T}_{i}=(-1)^{(p(D-p)+s)}T_{i}$ where $p$ is
the degree of the dual field strength and $s$ is the signature of
the spacetime.The degree of the dual field strengths corresponding
to the dual generators is $(D-1)$ in our case.In [3] the signature
of the spacetime is assumed to be $(D-1)$ for this reason
$\mathcal{S}$ sends the dual generators to the scalar ones with a
positive sign.In this work we will assume that the signature is
$s=1$.Thus the sign factor above is dependent on the spacetime
dimension $D$ and we have
$\mathcal{S}\widetilde{T}_{i}=(-1)^{D}T_{i}$.Now by following the
same steps in [1,2,3] and using the twisted self-duality equation
we can express the doubled field strength as

\begin{equation}
\mathcal{G}^{\prime}=d\nu\nu^{-1}+\frac{1}{2}(-1)^{D}\ast
d\phi^{i}\widetilde{H}_{i}+(-1)^{D}e^{\frac{1}{2}\alpha_{i}
\phi^{i}}\ast F^{\alpha}\widetilde{E}_{\alpha}.
\end{equation}

The Cartan form $\mathcal{G}_{0}=d\nu\nu^{-1}$ is already
calculated in (3.17).As explained in [3] the generators
$\{T_{i}\}$ are even and $\{\widetilde{T}_{i}\}$ are even or odd
depending on the spacetime dimension $D$ within the context of the
differential algebra generated by the solvable algebra generators,
their duals and the differential forms.By using the properties of
this differential algebra [2] and the fact that from it's
definition $\mathcal{G}^{\prime}$ obeys the Cartan-Maurer equation

\begin{equation}
d\mathcal{G}^{\prime }-\mathcal{G}^{\prime }\wedge
\mathcal{G}^{\prime }=0
\end{equation}

we can show that if we choose

\begin{gather}
[T_{i},\widetilde{H}_{j}]=0,\notag\\
\notag\\
[H_{j},\widetilde{E}_{\alpha }]=-\alpha _{j}\widetilde{E}_{\alpha }\quad ,\quad [%
E_{\alpha },\widetilde{E}_{\alpha }]=\frac{1}{4}\overset{r}{\underset{j=1}{%
\sum }}\alpha _{j}\widetilde{H}_{j},\notag\\
\notag\\
[E_{\alpha },\widetilde{E}_{\beta }]=N_{\alpha ,-\beta }\widetilde{E}%
_{\gamma },\quad\quad\alpha -\beta =-\gamma,\alpha \neq \beta.
\end{gather}

for $i=1,...,$dim$\mathbf{s}_{0}$,$j=1,...,r$ and
$\alpha,\beta.\gamma\in\Delta_{nc}^{+}$ then (4.4) by inserting
(4.3) gives the correct second order equations (3.19).As a matter
of fact if we choose in general
$[E_{\alpha},\widetilde{H}_{j}]=a_{\alpha
j}\alpha_{j}\widetilde{E}_{\alpha}$ and
$[H_{j},\widetilde{E}_{\alpha }]=b_{j\alpha}\alpha
_{j}\widetilde{E}_{\alpha }$ with $a_{j\alpha},b_{\alpha j}$
arbitrary but obeying the constraint $a_{j\alpha}+b_{\alpha j}=-1$
in addition to the rest of the commutators in (4.5) we can satisfy
the second order equations (3.19).However for simplicity like in
[3] we will choose $a_{j\alpha}=0$ as seen in (4.5).Therefore the
structure constants in (4.2) become

\begin{gather}
\widetilde{f}_{\alpha m}^{n}=0,\quad\quad m\leq r\quad,\quad
\widetilde{f}_{\alpha ,\alpha +r}^{i}=\frac{1}{4}\alpha _{i},\quad\quad%
i\leq r\notag\\
\notag\\
\widetilde{f}_{\alpha ,\alpha +r}^{i}=0,\quad\quad i>r\quad,\quad
\widetilde{f}_{\alpha ,\beta +r}^{i}=0,\quad\quad i\leq r,%
\alpha \neq \beta \notag\\
\notag\\
\widetilde{f}_{\alpha ,\beta +r}^{\gamma +r}=N_{\alpha ,-\beta },\quad\quad \alpha -\beta =-\gamma,\alpha \neq \beta\notag\\
\notag\\
\widetilde{f}_{\alpha ,\beta +r}^{\gamma +r}=0,\quad\quad \alpha
-\beta \neq -\gamma ,\alpha \neq \beta .
\end{gather}

Also

\begin{gather}
\widetilde{g}_{im}^{n}=0,\quad\quad m\leq r\quad,\quad\widetilde{%
g}_{im}^{n}=0,\quad\quad m>r,m\neq n\notag\\
\notag\\
\widetilde{g}_{i\alpha }^{\alpha }=-\alpha _{i},\quad\quad\alpha
>r.
\end{gather}

These relations with the commutation relations of the solvable Lie
algebra $\mathbf{s}_{0}$ form the complete algebraic structure of
the enlarged Lie superalgebra.In (4.3) $\mathcal{G}^{\prime}$ has
been given only in terms of the original scalar fields as the
twisted self-duality condition has been used primarily.From the
definition of $\nu^{\prime}(x)$ in (4.1) since we have obtained
the full set of commutation relations without using the twisted
self-duality condition we can explicitly calculate the Cartan form
$\mathcal{G}^{\prime }$ in terms of both the scalar fields and
their duals [3].

\begin{equation}
\begin{aligned}
\mathcal{G}^{\prime}&=d\nu^{\prime}\nu^{\prime-1}\\
\\
&=\frac{1}{2}d\phi ^{i}H_{i}+\overset{\rightharpoonup }{\mathbf{E}%
^{\prime }}\:\mathbf{\Omega }\:\overset{\rightharpoonup }{d\chi }+%
\overset{\rightharpoonup
}{\widetilde{\mathbf{T}}}\:e^{\mathbf{\Gamma
}}\:e^{\mathbf{\Lambda }}\:\overset{\rightharpoonup }{\mathbf{A}}.
\end{aligned}
\end{equation}

We have introduced the matrices
$(\Gamma)_{n}^{k}=\frac{1}{2}\phi^{i}\,\widetilde{g}_{in}^{k}$ and
$(\Lambda)_{n}^{k}=\chi^{m}\,\widetilde{f}_{mn}^{k}$.The row
vector $\overset{\rightharpoonup }{\widetilde{\mathbf{T}}}$ is
defined as $\{\widetilde{\mathbf{T}}^{i}\}$ and the column vector
$\overset{\rightharpoonup }{\mathbf{A}}$ is
$\mathbf{A}^{i}=\frac{1}{2}d\widetilde{\phi}^{i}$ for $i=1,...,r$
and
$\mathbf{A}^{\alpha+r}=d\widetilde{\chi}^{\alpha}$,$\alpha\in\Delta_{nc}^{+}$
in other words if we enumerate the roots in $\Delta_{nc}^{+}$ then
$\alpha=1,...,$dim$\mathbf{n}_{0}$.When we apply the twisted
self-duality condition $\ast \mathcal{G^{\prime}=SG^{\prime}}$ on
(4.8) we may achieve the first order equations locally whose
exterior derivative will give the second order equations (3.19),
[1,2].Therefore similar to the split case in [3] the locally
integrated first order field equations of the Lagrangian (3.15)
are

\begin{equation}
\ast \overset{\rightharpoonup }{\mathbf{\Psi }}=(-1)^{D}e^{\mathbf{\Gamma }%
}e^{\mathbf{\Lambda }}\overset{\rightharpoonup }{\mathbf{A}}.
\end{equation}

The column vector $\overset{\rightharpoonup }{\mathbf{\Psi }}$ is
defined as $\mathbf{\Psi}^{i}=\frac{1}{2}d\phi^{i}$ for $i=1,...r$
and
$\mathbf{\Psi}^{\alpha+r}=e^{\frac{1}{2}\alpha_{i}\phi^{i}}\mathbf{\Omega
}_{l}^{\alpha}d\chi^{l}$ where
$\alpha=1,...,$dim$\mathbf{n}_{0}$.Due to the assumed signature of
the spacetime these equations unlike the ones in [3] have a sign
factor depending on the spacetime dimension.We should also observe
that the case $SL(2,\mathbb{R)}$ /$SO(2)$ scalar coset of IIB
supergravity whose first-order equations are explicitly calculated
in [3] has a positive sign factor for both of the signatures
introduced here and in [3] since $D=10$.

Following the discussion in section three we can also find the
first order equations for the set $\{\varphi^{i}\}$.Firstly we can
define the transformed matrices
$\mathbf{\Gamma}^{\prime}(\varphi^{j})\equiv\mathbf{\Gamma}(\phi^{i}(\varphi^{j}))$
and
$\mathbf{\Lambda}^{\prime}(\varphi^{j})\equiv\mathbf{\Lambda}(\chi^{m}(\varphi^{j}))$
which can be obtained by calculating the local transformation
rules from (3.22) or (3.27).If we make the observation that the
right hand side of the differential form of the transformation
between $\{\varphi^{i}\}$ and $\{\phi^{i},\chi^{\alpha}\}$ namely
(3.27) are the components of $\overset{\rightharpoonup
}{\mathbf{\Psi }}$,from (4.9) we can write down the first order
equations for $\{\varphi^{i}\}$ as

\begin{equation}
\mathbf{\Delta}\ast \overset{\rightharpoonup }{\mathbf{d\varphi
}}=(-1)^{D}e^{\mathbf{\Gamma }^{\prime}}e^{\mathbf{\Lambda
}^{\prime}}\overset{\rightharpoonup }{\mathbf{A}}.
\end{equation}
We may also transform (4.10) so that we do not  have to calculate
the explicit transformations between the fields.Firstly we should
observe that the structure constants $\{\widetilde{g}_{in}^{k}\}$
and $\{\widetilde{f}_{mn}^{k}\}$ of (4.2) form a representation
for $\mathbf{s}_{0}$ as the representatives of the basis
$\{H_{i}\}$ and $\{E_{m}\}$ respectively.Thus under the
representation
\begin{equation}
\begin{aligned}
e^{\frac{1}{2}\phi ^{i}H_{i}}e^{\chi ^{m}E_{m}}&=e^{\mathbf{\Gamma
}}e^{\mathbf{\Lambda
}}\\
\\
&\equiv
e^{\frac{1}{2}\phi^{i}\,\widetilde{g}_{in}^{k}}e^{\chi^{m}\,\widetilde{f}_{mn}^{k}}.
\end{aligned}
\end{equation}

In section three we have assumed that locally

\begin{equation}
e^{\varphi^{i}T_{i}}=e^{\frac{1}{2}\phi ^{i}H_{i}}e^{\chi
^{m}E_{m}}.
\end{equation}
Therefore the first-order equations (4.10) for $\{\varphi_{i}\}$
can be written as
\begin{equation}
\mathbf{\Delta}\ast \overset{\rightharpoonup }{\mathbf{d\varphi
}}=(-1)^{D}e^{\mathbf{\Pi}}\overset{\rightharpoonup }{\mathbf{A}}
\end{equation}

where we have defined
$(\mathbf{\Pi})_{n}^{k}=\sum\limits_{i=1}^{r}\varphi^{i}\widetilde{g}_{in}^{k}+\sum\limits
_{m=r+1}^{dim\mathbf{n}_{0}}\varphi^{m}\widetilde{f}_{mn}^{k}$.by
using the representation established by (4.2).

One may also obtain these first order equations independently by
applying the dualisation method on the parametrization (2.3).We
again assume the solvable Lie algebra parametrization.Let us first
define the doubled coset map
\begin{equation}
\nu ^{\prime\prime }=e^{\varphi^{i}T_{i}}e^{\widetilde{\varphi
}^{i}\widetilde{T}_{i}}
\end{equation}
in which we have introduced the dual fields and generators as
usual.If we calculate the Cartan form
$\mathcal{G}^{\prime\prime}=d\nu ^{\prime\prime }\nu
^{\prime\prime-1}$ by carrying out similar calculations like we
have done before we find that

\begin{equation}
\begin{aligned}
\mathcal{G}^{\prime\prime}&=d
e^{\varphi^{i}T_{i}}e^{-\varphi^{i}T_{i}}+e^{\varphi^{i}T_{i}} d
e^{\widetilde{\varphi
}^{i}\widetilde{T}_{i}}e^{-\widetilde{\varphi
}^{i}\widetilde{T}_{i}}e^{-\varphi^{i}T_{i}}\\
\\
&=\overset{\rightharpoonup
}{\mathbf{T}}\:\mathbf{\Delta}\:\overset{\rightharpoonup
}{\mathbf{d\varphi}}+\overset{\rightharpoonup
}{\widetilde{\mathbf{T}}}\:e^{\mathbf{\Pi}}\:\overset{\rightharpoonup
}{\mathbf{d\widetilde{\varphi}}}.
\end{aligned}
\end{equation}

The first term has already been calculated in (3.25).We have
calculated the structure constants related to the dual generators
in (4.6) and (4.7).If we apply the twisted self-duality equation
$\ast \mathcal{G^{\prime\prime}=SG^{\prime\prime}}$ above we find
the first order equations as

\begin{equation}
\mathbf{\Delta}\ast \overset{\rightharpoonup }{\mathbf{d\varphi
}}=(-1)^{D}e^{\mathbf{\Pi}}\overset{\rightharpoonup
}{\mathbf{d\widetilde{\varphi}}}.
\end{equation}

Since the dual fields are auxiliary fields we can always choose
$(\mathbf{d\widetilde{\varphi}})^{i}=\mathbf{A}^{i}$ thus the
equations (4.13) and (4.16) are the same equations.This result
verifies the validity of (4.13) which is obtained by using the
transformation law (3.27) in (4.9).

Finally we should point out the fact that the case when the global
symmetry group $G$ is in split real form which is analyzed in
detail in [3] can be obtained by choosing $r=l$ (the rank of $G$)
and $\Delta_{nc}^{+}=\Delta^{+}$ in the expressions given in this
section.

\section{The $O(p,q)/(O(p)\times O(q))$ Scalar Cosets of the $T^{10-D}-$Compactified
Heterotic String Theory}

In this section we will briefly discuss the correspondence between
our results and the scalar cosets constructed in [4] by
identifying the solvable Lie algebra generators and by comparing
the coset parametrizations.The Kaluza-Klein compactification of
the Bosonic sector of the ten dimensional simple supergravity
which is coupled to $N$ Abelian gauge multiplets on the Euclidean
Tori $T^{10-D}$ is discussed in [4].The scalar sectors of the
resulting $D$-dimensional theories are formulated by the
introduction of the $G/K$ coset spaces.When as a special case the
number of U(1) gauge fields is chosen to be 16 the formulation
corresponds to the dimensional reduction of the low-energy
effective Bosonic Lagrangian of the ten dimensional heterotic
string theory.The global symmetries of the Bosonic sectors of
these reduced theories are also studied in detail in [4].

As it is clear from the mainline of [4] the scalar Lagrangian of
the $D$-dimensional compactified theories for $9\geq D\geq 3$ can
be described in the form of (3.15) with an additional dilatonic
kinetic term after certain field redefinitions.It is also shown
that the $G/K$ coset representative $\nu$ and the internal metric
$\mathcal{M}=\nu^{T}\nu$ are elements of $O(10-D,10-D+N)$.The
determination of the fiducial point $W_{0}=$diag$(1,1,...,1)$ by
choosing all the scalar fields in the coset representative $\nu$
zero enables the identification of the isotropy group as
$O(10-D)\times O(10-D+N)$.Therefore the scalar manifold for the
$D$-dimensional compactified theory with $N$ gauge multiplet
couplings becomes

\begin{gather}
\frac{O(10-D,10-D+N)}{O(10-D)\times O(10-D+N)}\times\mathbb{R}.
\end{gather}

The extra $\mathbb{R}$ factor arises since there is an additional
dilaton which is decoupled from the rest of the scalars in the
scalar Lagrangian.In [4] it is also shown that
$O(10-D,10-D+N)\times\mathbb{R}$ is the global symmetry of not
only the scalar Lagrangian but the entire $D$-dimensional Bosonic
Lagrangian aswell.Here again $\mathbb{R}$ corresponds to the
constant shift symmetry of the decoupled dilaton.Furthermore the
$D=4$ and the $D=3$ cases are studied separately in [4] since they
have symmetry enhancements in addition to the general scheme of
(5.1).When the two-form potential is dualised with an additional
axion in $D=4$,an axion-dilaton $SL(2,\mathbb{R})$ system [1,2,3]
which is decoupled from the rest of the scalars occurs in the
scalar Lagrangian and the enlarged $D=4$ scalar manifold becomes

\begin{gather}
\frac{O(6,6+N)}{O(6)\times
O(6+N)}\times\frac{SL(2,\mathbb{R})}{O(2)}.
\end{gather}

On the other hand in $D=3$ apart from the original scalars the
remaining Bosonic fields are dualised to give $7+7+N$ additional
axions so that the entire Bosonic Lagrangian is composed of only
the scalars.The $D=3$ scalar manifold then becomes

\begin{gather}
\frac{O(8,8+N)}{O(8)\times O(8+N)}.
\end{gather}

We see that all of the global symmetry groups in (5.1),(5.2),(5.3)
apart from the contributions of the decoupled scalars namely
$O(10-D,10-D+N),O(6,6+N),O(8,8+N)$ are real forms of a non-compact
semisimple Lie group and they enable solvable Lie algebra
parametrizations of the cosets generated by the denominator groups
$O(10-D)\times O(10-D+N),O(6)\times O(6+N),O(8)\times O(8+N)$
respectively.In fact the orthogonal algebras o(p,q) are elements
of the $D_{n}$-series when $p+q=2n$ and the $B_{n}$-series when
$p+q=2n+1$.Depending on the values of p and q the group $O(p,q)$
can be in split real form or not.Forexample $O(2,3)$ is in split
real form.In the direction of the observation mentioned above the
analysis of the previous sections can be applied for the scalar
cosets of (5.1),(5.2),(5.3).However we should inspect the
realizations of these cosets in [4] from a closer point of view.It
is shown in [4] that if one assigns the set of generators
$\{H^{i},E_{i}^{j},V^{ij},U_{I}^{i}\}$ to the scalar fields
$\{\phi_{i},A^{i}_{(0)j},A_{(0)ij},B^{I}_{(0)i}\}$ resulting from
the dimensional reduction in [4] respectively,the scalar
Lagrangian of the compactified $D$-dimensional theory apart from
the decoupled scalars,for all of the cases described above can be
constructed in the form of (3.15) by using the coset
parametrization

\begin{equation}
\nu=e^{\frac{1}{2}\phi_{i}H^{i}}e^{A^{i}_{(0)j}E_{i}^{j}}e^{\frac{1}{2}A_{(0)ij}
V^{ij}}e^{B^{I}_{(0)i}U_{I}^{i}}.
\end{equation}

The non-vanishing commutators of the generators are calculated in
order that they lead to the scalar Lagrangian in [4] as

\begin{gather}
[\overset{\rightharpoonup }{H},E_{i}^{j}]=\overset{\rightharpoonup
}{\mathbf{b}}_{ij}E_{i}^{j}\quad,\quad[\overset{\rightharpoonup
}{H},V^{ij}]=\overset{\rightharpoonup
}{\mathbf{a}}_{ij}V^{ij}\quad,\quad[\overset{\rightharpoonup
}{H},U_{I}^{i}]=\overset{\rightharpoonup
}{\mathbf{c}}_{i}U_{I}^{i},\notag\\
\notag\\
[E_{i}^{j},E_{k}^{l}]=\delta_{k}^{j}E_{i}^{l}-\delta_{i}^{l}E_{k}^{j}\quad,\quad
[E_{i}^{j},V^{kl}]=-\delta_{i}^{k}V^{jl}-\delta_{i}^{l}V^{kj},\notag\\
\notag\\
[E_{i}^{j},U_{I}^{k}]=-\delta_{i}^{k}U_{I}^{j}\quad,\quad
[U_{I}^{i},U_{J}^{j}]=\delta_{IJ}V^{ij}
\end{gather}

 where $\overset{\rightharpoonup
}{\mathbf{a}}_{ij},\overset{\rightharpoonup
}{\mathbf{b}}_{ij},\overset{\rightharpoonup }{\mathbf{c}}_{i}$ are
the dilaton vectors whose detailed descriptions can be found in
[4].We note that since in the $D$-dimensional
$T^{10-D}$-compactified theory the scalars are coupled to the
one-form potentials which form the ($20-2D+N$)-dimensional
fundamental representation of $O(10-D,10-D+N)$,both the coset
representative (5.4),the internal metric $\mathcal{M}$ and the
generators in (5.5) are represented by ($20-2D+N$)-dimensional
matrices.

In [4] an embedding of the algebra (5.5) into $o(10-D,10-D+N)$ is
also given as

\begin{gather}
H_{i}=(2)^{1/2}h_{\widetilde{e}_{i}}\quad,\quad
E_{i}^{j}=E_{-\widetilde{e}_{i}+\widetilde{e}_{j}} \quad,\quad
V^{ij}=E_{\widetilde{e}_{i}+\widetilde{e}_{j}}
,\notag\\
\notag\\
U_{2k-1}^{i}=(2)^{-1/2}(E_{\widetilde{e}_{i}+e_{k+m}}+E_{\widetilde{e}_{i}-e_{k+m}}),\notag\\
\notag\\
U_{2k}^{i}=i(2)^{-1/2}(E_{\widetilde{e}_{i}+e_{k+m}}-E_{\widetilde{e}_{i}-e_{k+m}})
\end{gather}

where $1\leq i<j\leq 10-D$.We have defined $m=10-D$ and $1\leq
k\leq[N/2]$.When $N$ is odd in addition to (5.6) we also have

\begin{gather}
U_{N}^{i}=E_{\widetilde{e}_{i}}.
\end{gather}

In (5.6) and (5.7) $\{e_{i}\}$ is an orthonormal basis with a
representation of null entries except the $i$'th position which is
equal to one.It is used to characterize the positive roots of
$o(10-D,10-D+N)$ [4] and $\{\widetilde{e}_{i}\}$ are defined as

\begin{gather}
\widetilde{e}_{i}=e_{11-D-i},\quad\quad 1\leq i\leq 10-D.
\end{gather}

As it is mentioned in [4] owing to their definitions the
generators $\{h_{\widetilde{e}_{i}}\}$ appearing in (5.6) form a
basis for $\mathbf{h}_{k}$ and the ones $\{E_{\widetilde{e}_{i}\pm
e_{j}},E_{\widetilde{e}_{i}\pm\widetilde{e}_{j}},E_{\widetilde{e}_{i}}\}$
appearing in (5.6) and (5.7) are the generators of
$\mathbf{n}_{0}$ (the last set $\{E_{\widetilde{e}_{i}}\}$ is
included when $N$ is odd and excluded when $N$ is even).Therefore
we conclude by following the discussion of section two that the
algebra structure presented in (5.5) corresponds to the solvable
Lie algebra of $o(10-D,10-D+N)$.We can identify the basis
$\{h_{\widetilde{e}_{i}},E_{\widetilde{e}_{i}\pm
e_{j}},E_{\widetilde{e}_{i}\pm\widetilde{e}_{j}},E_{\widetilde{e}_{i}}\}$
with the one $\{H_{i},E_{\alpha}\}$ which we have used in our
previous analysis in sections three and four bearing in mind that
the even $N$ and the odd $N$ cases are differing.

We observe that (5.6) and (5.7) give us the transformations
between the abstract solvable Lie algebra generators we have used
in the previous sections and the original generators which arise
in the coset formulations of the $D$-dimensional compactified
theories of [4].Therefore the embedding of the generators of (5.4)
and (5.5) can be used to derive the exact and the differential
form of the transformation between the associated scalar fields
defined in the coset parametrizations (5.4) and (3.5).Furthermore
we should also remark that since the exponential factors are not
the same in (3.5) and (5.4) by using a similar analysis which is
held in the last part of section three one needs to find the local
transformations of these different parametrizations.Thus one has
to express the coset (5.4) in terms of
$\{h_{\widetilde{e}_{i}},E_{\widetilde{e}_{i}\pm
e_{j}},E_{\widetilde{e}_{i}\pm\widetilde{e}_{j}},E_{\widetilde{e}_{i}}\}$
by using (5.6) and (5.7) then search for a local transformation
law to write it as (3.5).Refering to section four we need only to
determine the structure constants of the solvable Lie algebra of
$o(10-D,10-D+N)$ generated by the generators
$\{h_{\widetilde{e}_{i}},E_{\widetilde{e}_{i}\pm
e_{j}},E_{\widetilde{e}_{i}\pm\widetilde{e}_{j}},E_{\widetilde{e}_{i}}\}$
(the last set $\{E_{\widetilde{e}_{i}}\}$ is included when N is
odd) to find the first-order equations.We see that we do not need
to use the ($20-2D+N$)-dimensional fundamental representatives of
the generators $\{H^{i},E_{i}^{j},V^{ij},U_{I}^{i}\}$ [4] or the
ones of $\{h_{\widetilde{e}_{i}},E_{\widetilde{e}_{i}\pm
e_{j}},E_{\widetilde{e}_{i}\pm\widetilde{e}_{j}},E_{\widetilde{e}_{i}}\}$
to find the first-order equations.One would certainly need the
fundamental representation when one considers the gauge multiplet
coupling.Finally the transformation laws would enable us to
express the first-order formulations of the scalar cosets
constructed in [4] in terms of the original scalar fields which
the dimensional reduction contributes.

\section{Conclusion}
The solution space of the scalar sector of a wide class of
supergravity theories is generated by the $G/K$ symmetric space
sigma model.After introducing the algebraic structure and the
suitable parametrizations for the coset $G/K$ in section two we
have given two equivalent formulations of the non-split coset
sigma model and studied them in detail in section three.The
solvable Lie algebra parametrization is used to derive the field
equations for both of these formulations.In general we have formed
an analysis on two different coset maps which define two sets of
scalar fields.A local transformation between these two different
sets is also constructed in section three.In section four we have
generalized the results of [3] for the case when the non-compact
symmetry group $G$ is not necessarily in split real form.From the
choice of the solvable Lie algebra parametrization given in
section two it is apparent that the split case whose dualisation
is introduced in [3] for a different spacetime signature
convention than the one assumed here is a special example of the
general formulation constructed in this note.Thus the formulation
given here contains the split case as a limiting example in the
group theory sense.We have derived the local first-order equations
for a generic non-split coset by using the coset map of [1,2].We
have also obtained the first order equations for the scalar fields
which are defined through a more conventional coset map by using
the explicit transformation defined in section three and by
applying a separate dualisation on the later coset map.

In section five we have discussed a possible field of application
of our results namely we have presented the link between the
abstract formulation of the symmetric space sigma model in section
three and four and the scalar coset realizations of [4] which
arise in the Kaluza-Klein reduction of the ten dimensional simple
supergravity coupled to $N$ Abelian gauge multiplets on
$T^{10-D}$.In particular $N=16$ case corresponds to the
$T^{10-D}$-compactification of the ten-dimensional low-energy
effective heterotic string theory.A transformation between the
scalar field definitions given in section three and the original
scalar fields used in [4] can be explored by using the
transformation of the generators given in section five and by
considering the coset parametrizations.This would enable a direct
construction of the first-order formulation of the scalar cosets
presented in [4].

The dualisation and the local first order formulation of section
four is an extension of [3].This work improves the application of
the dualisation method of [1,2] from the split-coset case of [3]
to the entire set of non-compact symmetric space scalar cosets of
supergravities.The results presented here are powerful since they
would be the starting point and an essential part of the first
order formulations of the pure and the matter coupled
supergravities which are not studied in [1,2].

 In summary besides studying the field equations and different
parametrizations of the non-split
 coset,this work completes the dual formulation of the symmetric
space
 sigma model when the rigid symmetry group is a real form of a
 non-compact semisimple Lie group.It extends the construction
 in [3] which is performed for the special split case.

\section{Acknowledgements}
This work has been supported by TUBITAK (The Scientific and
Technical Research Council of Turkey).I would like to thank Prof
Tekin Dereli and Prof Peter West for discussions,motivation and
useful remarks also Dr Arjan Keurentjes for motivation and
explanations.

\end{document}